# OWL: Yet to arrive on the Web of Data?

Birte Glimm
Ulm University, Institute of Artificial Intelligence, 89069 Ulm, Germany

Aidan Hogan
Digital Enterprise Research Institute, National University of Ireland Galway, Ireland

Markus Krötzsch
University of Oxford, Department of Computer Science, OX1 3QD Oxford, United Kingdom

Axel Polleres
Siemens AG Österreich, Siemensstrasse 90, 1210 Vienna, Austria

## ABSTRACT

Seven years on from OWL becoming a W3C recommendation, and two years on from the more recent OWL 2 W3C recommendation, OWL has still experienced only patchy uptake on the Web. Although certain OWL features (like owl:sameAs) are very popular, other features of OWL are largely neglected by publishers in the Linked Data world. This may suggest that despite the promise of easy implementations and the proposal of tractable profiles suggested in OWL's second version, there is still no "right" standard fragment for the Linked Data community. In this paper, we (1) analyse uptake of OWL on the Web of Data, (2) gain insights into the OWL fragment that is actually used/usable on the Web, where we arrive at the conclusion that this fragment is likely to be a simplified profile based on OWL RL, (3) propose and discuss such a new fragment, which we call OWL LD (for Linked Data).

## 1. INTRODUCTION

Under the initial impetus of the Linking Open Data project – and guided by the Linked Data principles [3] and associated best-practices – a rich vein of openly-available structured data has been published on the Web using Semantic Web standards. Publishing RDF on the Web is no longer confined to academia and hobbyists: the current "Web of Data" now features exports from various corporate and commercial bodies (e.g., BBC, New York Times, Freebase, BestBuy), online communities (e.g., Wikipedia, Geonames), life-science corpora (e.g., DrugBank, Linked Clinical Trials) and governmental bodies (e.g., data.gov, data.gov.uk, EuroStat). The "Linked Open Data cloud" now depicts 295 interlinked datasets, which together consist of an estimated 31.6 billion RDF triples.[1]

Although RDF provides standard syntaxes and a common data-model for disseminating structured information, it offers very little when it comes to giving semantics to the published data. RDF Schema (RDFS) and OWL were developed to address this by providing a vocabulary for describing schema data. The special vocabulary terms of RDFS and OWL – such as rdfs:subClassOf or owl:FunctionalProperty – have a well-defined semantics, which can be used to derive implicit consequences from the data.

In terms of publishing, parts of the RDFS and OWL standards have been adopted on the Web of Data. Linked Data literature recommends use of owl:sameAs relations to denote when two URIs refer to the same resource [18, § 2.5.2]. Further, Linked Data guidelines recommend use of RDFS [18, § 4.4.2] for defining terms and interlinking vocabularies. As regards the broader OWL standard, current guidelines explicitly mention use of owl:equivalentClass, owl:equivalentProperty, owl:InverseFunctionalProperty and owl:inverseOf [18, § 4.4.2]. However, other OWL features are not mentioned.

In terms of standards, RDFS and OWL 1 pre-date the Linked Data movement and are not directly tailored towards Linked Data requirements. Although the informative entailment rules for supporting RDFS inferences are *relatively* straightforward, things like the infinitely many entailed axiomatic triples reduce its practicality [28]. In OWL 1 the situation is more complex: OWL 1 Full further extends the RDFS semantics to the extent that reasoning becomes undecidable. In OWL 1 DL and OWL 1 Lite, where the semantics are based on Description Logics, typical reasoning tasks remain decidable, but are of exponential or harder worst-case complexity. OWL 2 addresses the complexity issue by defining *profiles* [6]: fragments for which at least some reasoning tasks are tractable. Reasoning with inconsistent data is, however, still problematic in any OWL fragment. Further, each profile is a syntactic subset of OWL DL such that RDF data must adhere to certain non-trivial conditions which are commonly not followed in Web ontologies [2, 38, 7]. However, OWL RL includes a ruleset called OWL RL/RDF, which is applicable over arbitrary RDF data.

Although the OWL RL profile is implementable using straight-forward rule-based technologies, (as we will see) the profile still includes many features with sparse uptake in Linked Data vocabularies. Which features are prominently used is, however, unclear. In order to clarify this, we survey a broad spectrum of RDF Web data and measure uptake of individual RDFS and OWL features used therein. Since datatypes also play a role for OWL reasoning, we additionally look at the use of datatypes in published data. We further analyse to what extent OWL features are supported by tools that provide the technical infrastructure for building complex Semantic Web applications.

Our analysis suggests that a much simpler profile of OWL might be better targeted towards the current needs of the Linked Data community. We thus propose OWL LD (for Linked Data) as a subset of the OWL RL profile, using the insights of our survey to make an informed decision as to which features of the RDFS and OWL standards should be included in the profile.

The remainder of the paper is structured as follows: In the next section, we introduce some preliminaries. In Section 3, we present our survey of the use of RDFS and OWL features on the Web, in-

---

[1] http://www4.wiwiss.fu-berlin.de/lodcloud/state/

This work has been funded in part by Science Foundation Ireland under Grant No. SFI/08/CE/I1380 (Líon-2) and by an IRCSET postgraduate grant.
*CoRR* Copyright 2012.

cluding a survey of datatypes. In Section 4, we analyse the tool support for RDFS and OWL. Drawing upon our observations, we propose and define the OWL LD profile in Section 5, and discuss formal aspects of reasoning over the profile in Section 6. Next, in Section 7 we give a synopsis of related work for empirical analyses of RDFS and OWL data on the Web. We conclude in Section 8.

## 2. BACKGROUND

Before analysing the use of OWL in the web, we first recall some relevant features of RDF, RDFS, and OWL semantics and give a summary of the existing OWL profiles.

### 2.1 RDF Graphs and Their Semantics

Given the set of URI references $\mathsf{U}$, the set of blank nodes $\mathsf{B}$, and the set of literals $\mathsf{L}$, the set of *RDF constants* is denoted by $\mathsf{C} := \mathsf{U} \cup \mathsf{B} \cup \mathsf{L}$. We use CURIEs to denote URIs (e.g., owl:sameAs), where the prefixes used in this paper can be looked up, e.g., at http://prefix.cc/. We often use Turtle syntax; e.g., we may use a as a shortcut for rdf:type. Finally, $\mathsf{V}$ denotes the set of *RDF variables* ranging over $\mathsf{C}$ and we prefix variables with '?'.

An *RDF triple* $(s, p, o)$ is a triple from the set of all *RDF triples* $\mathsf{G} := \mathsf{U} \cup \mathsf{B} \times \mathsf{U} \times \mathsf{C}$, where $s$ is called subject, $p$ predicate, and $o$ object. We call a finite set of triples $G \subset \mathsf{G}$ an *RDF graph*.

Semantically, RDF graphs can be interpreted in a number of ways based on various W3C recommendations. The *simple semantics* [17] considers only the graph structure of RDF, whereas more elaborate semantics such as RDFS entailment [17] or the OWL 2 Direct and RDF-Based Semantics (see below) provide special meanings for certain terms.

The common basis for all such semantics is that they are specified in terms of model theory: one defines *interpretations* together with necessary and sufficient conditions that specify when an interpretation *satisfies* a graph. When defining a semantics $\mathsf{E}$ (such as RDF, RDFS, etc.) one often speaks of $\mathsf{E}$-interpretations and $\mathsf{E}$-satisfaction. The set of all $\mathsf{E}$-interpretations that $\mathsf{E}$-satisfy a graph $G$ are called the $\mathsf{E}$-*models* of $G$. Semantic entailment follows from this notion: a graph $G$ $\mathsf{E}$-*entails* a graph $G'$, written $G \models_\mathsf{E} G'$, if and only if every $\mathsf{E}$-model of $G$ is also an $\mathsf{E}$-model of $G'$.

### 2.2 OWL and its Fragments

OWL 2 is an ontology language that provides advanced schema modelling capabilities that can be used together with RDF data. OWL 2 supersedes the earlier specification "OWL 1" by introducing new modelling features, additional serialisations, updated conformance conditions and various corrections. When omitting the version number, we thus mean the current standard OWL 2.

Every RDF graph can be considered as an OWL ontology and the language of all RDF documents is called *OWL Full* to emphasise that all such graphs should be viewed as ontologies. In applications, however, OWL ontologies are usually viewed as being composed of *axioms*, that can be more complex than single triples. For example, the triple ex:a owl:sameAs ex:b . corresponds to the OWL axiom SameIndividual(ex:a ex:b) whereas the axiom

$$\text{ObjectPropertyRange(skos:member} \atop \text{ObjectUnionOf(skos:Concept skos:Container))} \quad (1)$$

expands to the six RDF triples

$$\text{skos:member rdfs:range \_:x. \_:x owl:unionOf \_:x1 .} \atop \text{\_:x1 rdf:first skos:Concept. \_:x1 rdf:rest \_:x2 .} \quad (2) \atop \text{\_:x2 rdf:first skos:Container. \_:x2 rdf:rest rdf:nil .}$$

Additional conditions need to be imposed on RDF graphs to ensure that they are in one-to-one correspondence to a collection of OWL axioms. A syntactic subset of OWL Full for which this is possible is *OWL DL*, which also imposes further restrictions that are useful for computing semantic conclusions from the ontology [27]. It can still be computationally expensive to draw conclusions from OWL DL ontologies. To address this, OWL further defines three syntactically restricted sub-languages (*profiles*) of OWL DL called *OWL EL*, *OWL RL* and *OWL QL* [6] (see also Table 2 later for a brief feature comparison). OWL Full, OWL DL and the OWL profiles together constitute the five language fragments of OWL. The essential features of RDF Schema (sub-classes and -properties, domain, range) are covered by all of these fragments, but only OWL Full supports arbitrary RDF documents.

Various further sub-languages of OWL have been proposed outside of the official standard. The current profiles themselves have been inspired by existing approaches: $\mathcal{EL}^{++}$ for OWL EL [24], DL-Lite [5] for OWL QL, and Description Logic Programs (DLP) [13] and pD* [35] for OWL RL. Generally, these approaches aimed to maximise the expressivity and thus approach the current standard quite closely (albeit, only for OWL 1 features). DLP is defined as a syntactic fragment of OWL. Other languages – including pD* – came about by extending RDFS with some additional features. Allemang and Hendler proposed *RDFS-Plus* based on an informal survey of practitioners and three criteria felt important for adoption: *pedagogism* (intuitive and easy to learn), *practicality* (real use-cases in modelling), and *computational feasibility* (not too hard to implement) [1]. This language was later extended to *RDFS 3.0* along similar principles [19]. Fisher et al. propose a similar profile to RDFS-Plus called *L2*, where the rationale for including or excluding features is given on an ad-hoc basis [11]. A more detailed overview of the main features for these languages is also found in Table 2.

### 2.3 OWL Semantics and Reasoning

OWL ontologies can be interpreted under two different semantics that agree in important cases: the *RDF-Based Semantics* (RS) [17] and the *Direct Semantics* (DS) [26]. Like in RDF(S), the semantics are defined by specifying a model theory, i.e., by defining valid interpretations for ontologies based on semantic conditions. In RS, these models are based on the representation of OWL axioms as RDF graphs and thus can be viewed as a refined form of RDF interpretation. In DS, models are directly defined based on the structure of OWL axioms in the conceptual framework of Description Logics (which in turn is based on first-order logic). Due to this, DS is only defined for ontologies that belong to the OWL DL language (or to any of its profiles) while RS can also be used on OWL Full. Besides this restriction, OWL language fragments are not tied to either semantics, leaving 9 valid combinations of syntactic fragment and formal semantics [34].

RS is arguably more robust since it is defined for any RDF graph while DS only works for ontologies in OWL DL. However, RS entailment (of derived facts) is undecidable so that concrete implementations can compute only a subset of the conclusions that the semantics specifies. In contrast, there are complete implementations for computing entailments under DS, though with a high (super-exponential) worst-case complexity if all of OWL DL is to be covered. When further restricting to the OWL profiles, entailment checking under DS can even be done in polynomial time. For RS, it is not known in general if the entailment problem becomes simpler in these cases. It is known, however, that RS and DS yield the same entailments on OWL RL under certain additional conditions, leading to a partial tractability result for RS for this particular case [6]. Similar results could be obtained in other cases since DS reasoning algorithms can typically be modified to obtain correct

(though often incomplete) RS reasoners.

DS reasoning in all of the OWL profiles and significant parts of OWL DL can be implemented using rules in a forward-chaining manner. For OWL RL, an algorithm is suggested in the specification [6], while other works have covered OWL EL [24] and parts of OWL DL that also cover OWL QL [33]. For OWL QL, query rewriting is a more common reasoning technique [5, 31]. There are many different reasoning techniques for OWL DL under DS, though not all of them lead to polynomial algorithms when applied to the OWL profiles. Two (necessarily incomplete) reasoning methods are known for RS: algorithms based on sets of derivation rules like the ones for OWL RL and an approach based on using first-order theorem provers [32].

## 3. SURVEY OF RDFS & OWL ADOPTION ON THE WEB OF DATA

We now present an empirical survey of RDFS & OWL adoption on the Web of Data. Our survey is conducted over the Billion Triple Challenge 2011 corpus, which consists of 2.145 billion quadruples crawled from 7.411 million RDF/XML documents through an open crawl ran in May/June 2011 spanning 791 pay-level domains. (A pay-level domain is a direct sub-domain of a top-level domain (TLD) or a second-level country domain (ccSLD), e.g., dbpedia.org, bbc.co.uk. This gives us our notion of "domain"). This corpus represents a broad *sample* of the Web of Data.

### 3.1 Measures Used

In order to adequately characterise the uptake of various RDF(S) and OWL features used in this corpus, we present different measures to quantify their *prevalence* and *prominence*.

First, we look at the *prevalence* of use of different features, i.e., how often they are used. Here, we must take into account the diversity of the data under analysis, where few domains account for a great many triples and many domains account for few triples, where certain domains tend to publish many small documents and others publish few large documents, and so forth [20]. We thus present three statistics: (1) number of *axioms* using the feature [Ax], (2) number of *documents* [Doc] and (3) number of *domains* [Dom].

Second, we look at the *prominence* of use of different features. We use PageRank to quantify our notion of prominence: PageRank calculates a variant of the Eigenvector centrality of nodes (e.g., documents) in a graph, where taking the intuition of directed links as "positive votes", the resulting scores help characterise the relative prominence of particular nodes on the Web [29, 15].

In particular, we first rank documents in the corpus. To construct the graph, we consider RDF documents as nodes, where a directed edge $(d_1, d_2)$ is extended from document $d_1$ to $d_2$ iff $d_1$ hosts RDF data that contains (in any triple position) a URI that dereferences to document $d_2$. This notion of dereferenceable links is core to Linked Data principles [3]. Note also that we follow redirects when checking dereferenceability. We then apply a standard PageRank analysis over the resulting directed graph, using the power iteration method with ten iterations. For reasons of space, we refer the interested reader to [29] for more detail on PageRank, and [20] for more detail on the particular algorithms used in this paper.

Given these rank scores, for the different RDF(S) and OWL features we then present (1) the *sum of PageRank scores* for documents in which they are used [∑ Rank]; (2) the *max PageRank score* of the highest-ranked document in which it appears [max Rank]; (3) the *max PageRank position* of that document in the ordering of the 7.411 million documents [max Pos].

In terms of intuition under the random surfer model of Page-

| # | Document URI | Rank |
|---|---|---|
| 1 | http://www.w3.org/1999/02/22-rdf-syntax-ns | 0.121 |
| 2 | http://www.w3.org/2000/01/rdf-schema | 0.110 |
| 3 | http://dublincore.org/2010/10/11/dcelements.rdf | 0.096 |
| 4 | http://www.w3.org/2002/07/owl | 0.078 |
| 5 | http://www.w3.org/2000/01/rdf-schema-more | 0.049 |
| 6 | http://dublincore.org/2010/10/11/dcterms.rdf | 0.036 |
| 7 | http://www.w3.org/2009/08/skos-reference/skos.rdf | 0.026 |
| 8 | http://xmlns.com/foaf/spec/ | 0.023 |
| 9 | http://dublincore.org/DCMI.rdf | 0.021 |
| 10 | http://www.w3.org/2003/g/data-view | 0.017 |
| 14 | http://id.loc.gov/authorities/sh98002267 | 4.01E-3 |
| 30 | http://motools.sourceforge.net/doc/musicontology.rdfs | 2.38E-3 |
| 38 | http://www.w3.org/.../wn20/schemas/wnfull.rdfs | 7.79E-4 |
| 43 | http://vivoweb.org/files/vivo-core-public-1.2.owl | 6.11E-4 |
| 87 | http://www.w3.org/2006/time | 2.07E-4 |
| 116 | http://rdf.geospecies.org/ont/geospecies | 1.22E-4 |
| 129 | http://motools.sourceforge.net/timeline/timeline.rdf | 1.06E-4 |
| 159 | http://vocab.org/bio/0.1/termgroup2.rdf | 8.11E-5 |
| 259 | http://www.ordnancesurvey.co.uk/.../geometry.owl | 4.39E-5 |
| 289 | http://www.ordnancesurvey.co.uk/.../admingeo.owl | 4.01E-5 |
| 990 | http://www.ordnancesurvey.co.uk/.../spatialrelations.owl | 1.24E-5 |

**Table 1: Top ten ranked documents and notable ranks (position < 1,000) mentioned later in Table 2**

Rank [29], given an agent starting from a random location and traversing documents on (our sample of) the Web of Data through randomly selected dereferenceable URIs, the ∑ Rank value for a feature approximately indicates the probability with which that agent will be at a document using that feature after traversing ten links. In other words, the score indicates the likelihood of an agent, operating over the Web of Data based on dereferenceable principles, to encounter a given feature.

The graph extracted from the corpus consists of 7.411 million nodes and 198.6 million edges. Table 1 presents the top ten ranked documents in our corpus, which are dominated by core meta-vocabularies, documents linked therefrom, and other popular vocabularies; we also present the ranks of other notable documents mentioned in the following section.[2]

### 3.2 Survey of RDF(S)/OWL Features

Table 2 presents the results of the survey of RDF(S) and OWL usage in our corpus, where for features with non-trivial semantics, we present the measures mentioned in the previous section, as well as support for the features in the different reasoning profiles discussed in Section 2.2. We exclude rdf:type, which appeared in 90.3% of documents. We present the table ordered by the sum of PageRank measure [∑ Rank]; recall that Table 1 provides a legend for notable documents (Pos<1,000).

In column BF, we indicate which features have expressions that can be represented as a single RDF triple, i.e., which features do not require auxiliary blank nodes of the form _:x or the SEQ production in Table 1 of the OWL 2 Mapping to RDF document [30]. This distinction is motivated by our initial observations that such features are typically the most widely used in Web data.

Figure 1 gives a visual impression of the sum of PageRank measure for the listed features (log scale), where different shades of grey are used to indicate to which vocabulary a term belongs (e.g., distinguishing the terms new in OWL 2 from the ones already in OWL 1).

Regarding *prevalence*, we see from Table 2 that owl:sameAs is the most widely used axiom in terms of documents (1.778 mil-

---
[2] We ran another similar analysis with links to and from core RDF(S) and OWL vocabularies disabled. The results for the feature analysis remained similar. Mainly owl:sameAs dropped several positions in terms of the sum of PageRank.

| # | Primitive | ∑ Rank | max Rank | max Pos | Ax | Doc | Dom | RDFS | L2 | RDFS+ | DLP | pD* | EL | QL | RL | BF |
|---|---|---|---|---|---|---|---|---|---|---|---|---|---|---|---|---|
| 1 | rdf:Property | 5.74E-1 | 1.21E-1 | 1 | 17,509 | 8,049 | 48 | ✓ | - | - | - | ✓ | - | - | - | ✓ |
| 2 | rdfs:range | 4.67E-1 | 1.21E-1 | 1 | 51,540 | 44,492 | 89 | ✓ | ✓ | ✓ | ✓ | ✓ | ✓ | ✓ | ✓ | ✓ |
| 3 | rdfs:domain | 4.62E-1 | 1.21E-1 | 1 | 97,288 | 43,247 | 89 | ✓ | ✓ | ✓ | ✓ | ✓ | ✓ | ✓ | ✓ | ✓ |
| 4 | rdfs:subClassOf | 4.60E-1 | 1.21E-1 | 1 | 1,164,620 | 115,608 | 109 | ✓ | ✓ | ✓ | ✓ | ✓ | ✓ | ✓ | ✓ | ✓ |
| 5 | rdfs:Class | 4.45E-1 | 1.21E-1 | 1 | 39,606 | 19,904 | 43 | ✓ | - | - | ✓ | ✓ | - | - | - | ✓ |
| 6 | rdfs:subPropertyOf | 2.35E-1 | 1.10E-1 | 2 | 11,490 | 6,080 | 80 | ✓ | ✓ | ✓ | ✓ | ✓ | ✓ | ✓ | ✓ | ✓ |
| 7 | owl:Class | 1.74E-1 | 7.80E-2 | 4 | 255,002 | 302,701 | 111 | - | - | ✓ | ✓ | ✓ | ✓ | ✓ | ✓ | ✓ |
| 8 | owl:ObjectProperty | 1.68E-1 | 7.80E-2 | 4 | 35,065 | 285,412 | 92 | - | - | ✓ | ✓ | ✓ | ✓ | ✓ | ✓ | ✓ |
| 9 | rdfs:Datatype | 1.68E-1 | 1.21E-1 | 1 | 31 | 23 | 9 | ✓* | - | - | ✓* | ✓* | ✓* | ✓* | ✓* | ✓* |
| 10 | owl:DatatypeProperty | 1.65E-1 | 7.80E-2 | 4 | 23,888 | 234,483 | 82 | - | - | ✓ | ✓ | - | ✓ | ✓ | ✓ | ✓ |
| 11 | owl:AnnotationProperty | 1.60E-1 | 7.80E-2 | 4 | 216 | 172,290 | 55 | - | - | - | ✓ | ✓ | ✓ | ✓ | ✓ | ✓ |
| 12 | owl:FunctionalProperty | 9.18E-2 | 2.63E-2 | 7 | 3,222 | 298 | 34 | - | - | ✓ | ✓ | ✓ | - | - | ✓ | ✓ |
| 13 | owl:equivalentProperty | 8.54E-2 | 3.57E-2 | 6 | 168 | 141 | 23 | - | - | ✓ | ✓ | ✓ | ✓ | ✓ | ✓ | ✓ |
| 14 | owl:inverseOf | 7.91E-2 | 2.63E-2 | 7 | 1,160 | 366 | 43 | - | ✓ | ✓ | ✓ | ✓ | - | ✓ | ✓ | ✓ |
| 15 | owl:disjointWith | 7.65E-2 | 2.63E-2 | 7 | 3,266 | 230 | 27 | - | - | - | ✓ | - | ✓ | ✓ | ✓ | ✓ |
| 16 | owl:sameAs | 7.29E-2 | 4.01E-3 | 14 | 3,450,554 | 1,778,208 | 117 | - | ✓ | ✓ | ✓ | ✓ | ✓ | - | ✓ | ✓ |
| 17 | owl:equivalentClass | 5.24E-2 | 2.32E-2 | 8 | 25,827 | 22,291 | 39 | - | ✓ | ✓ | ✓ | ✓ | ✓ | ✓ | ✓ | ✓ |
| 18 | owl:InverseFunctionalProperty | 4.79E-2 | 2.32E-2 | 8 | 75 | 111 | 24 | - | - | ✓ | ✓* | ✓ | - | - | ✓ | ✓ |
| 19 | owl:unionOf | 3.15E-2 | 2.63E-2 | 7 | 46,721 | 15,162 | 30 | - | - | - | ✓* | - | - | - | ✓* | - |
| 20 | owl:SymmetricProperty | 3.13E-2 | 2.63E-2 | 7 | 175 | 120 | 23 | - | ✓ | ✓ | ✓ | ✓ | - | ✓ | ✓ | ✓ |
| 21 | owl:TransitiveProperty | 2.98E-2 | 2.63E-2 | 7 | 223 | 150 | 30 | - | ✓ | ✓ | ✓ | ✓ | ✓ | ✓ | ✓ | ✓ |
| 22 | owl:someValuesFrom | 2.13E-2 | 1.65E-2 | 10 | 3,854 | 1,753 | 15 | - | - | - | ✓* | ✓* | ✓ | ✓* | ✓* | - |
| 23 | rdf:_* | 1.42E-2 | 8.11E-5 | 159 | 7,791,545 | 293,022 | 62 | ✓ | - | - | - | ✓ | - | - | - | - |
| 24 | owl:allValuesFrom | 2.98E-3 | 7.79E-4 | 38 | 108,989 | 29,084 | 20 | - | - | - | ✓* | ✓* | - | - | ✓* | - |
| 25 | owl:minCardinality | 2.43E-3 | 6.11E-4 | 43 | 395,841 | 33,309 | 19 | - | - | - | ✓* | - | - | - | - | - |
| 26 | owl:maxCardinality | 2.14E-3 | 6.11E-4 | 43 | 223,994 | 10,413 | 24 | - | - | - | ✓* | - | - | - | ✓* | - |
| 27 | owl:cardinality | 1.75E-3 | 7.79E-4 | 38 | 20,781 | 3,170 | 24 | - | - | - | ✓* | - | - | - | - | - |
| 28 | owl:oneOf | 4.13E-4 | 2.07E-4 | 87 | 736 | 74 | 11 | - | - | - | ✓* | - | ✓* | - | ✓* | - |
| 29 | owl:hasValue | 3.91E-4 | 2.07E-4 | 87 | 1,624 | 55 | 14 | - | - | - | ✓* | ✓ | ✓ | - | ✓ | - |
| 30 | owl:intersectionOf | 3.37E-4 | 1.06E-4 | 129 | 2,324 | 186 | 13 | - | - | - | ✓ | - | ✓ | ✓* | ✓ | - |
| 31 | owl:NamedIndividual[(2)] | 1.63E-4 | 1.22E-4 | 116 | 205 | 3 | 2 | - | - | - | - | - | ✓ | ✓ | ✓ | ✓ |
| 32 | owl:AllDifferent | 1.55E-4 | 1.22E-4 | 116 | 87 | 21 | 8 | - | - | - | - | - | ✓ | - | ✓ | ✓ |
| 33 | owl:propertyChainAxiom[(2)] | 1.23E-4 | 4.01E-5 | 289 | 52 | 14 | 6 | - | - | - | - | - | ✓ | - | ✓ | - |
| 34 | owl:onDataRange | 8.41E-5 | 4.39E-5 | 259 | 89 | 3 | 1 | - | - | - | - | - | - | - | - | - |
| 35 | owl:minQualifiedCardinality[(2)] | 8.40E-5 | 4.39E-5 | 259 | 7 | 2 | 1 | - | - | - | - | - | - | - | - | - |
| 36 | owl:qualifiedCardinality[(2)] | 4.02E-5 | 4.01E-5 | 289 | 95 | 2 | 1 | - | - | - | - | - | - | - | - | - |
| 37 | owl:AllDisjointClasses[(2)] | 4.01E-5 | 4.01E-5 | 289 | 9 | 2 | 2 | - | - | - | - | - | ✓ | ✓ | ✓ | - |
| 38 | owl:maxQualifiedCardinality[(2)] | 4.01E-5 | 4.01E-5 | 289 | 1 | 1 | 1 | - | - | - | - | - | - | - | ✓* | - |
| 39 | owl:ReflexiveProperty[(2)] | 1.30E-5 | 1.24E-5 | 990 | 1 | 2 | 1 | - | - | - | - | - | ✓ | ✓ | - | ✓ |
| 40 | owl:complementOf | 1.96E-6 | 6.28E-8 | 549,258 | 759 | 75 | 4 | - | - | - | ✓* | - | - | ✓* | ✓* | - |
| 41 | owl:differentFrom | 7.18E-7 | 6.81E-8 | 486,354 | 691 | 25 | 7 | - | - | ✓ | ✓ | ✓ | ✓ | - | ✓ | ✓ |
| 42 | owl:onDatatype | 2.72E-7 | 2.72E-7 | 70,414 | 2 | 1 | 1 | - | - | - | - | - | - | - | - | - |
| 43 | owl:disjointUnionOf | 6.31E-8 | 4.28E-8 | 1,005,307 | 2 | 2 | 2 | - | - | - | - | - | - | - | - | - |
| 44 | owl:hasKey[(2)] | 3.67E-8 | 3.67E-8 | 1,336,720 | 1 | 1 | 1 | - | - | - | - | - | ✓ | - | ✓ | - |
| 45 | owl:propertyDisjointWith[(2)] | 2.43E-8 | 2.43E-8 | 3,911,874 | 4 | 1 | 1 | - | - | - | - | - | - | ✓ | ✓ | ✓ |

Not Used: rdfs:ContainerMembershipProperty, owl:AllDisjointProperties[(2)], owl:Annotation[(2)], owl:AsymmetricProperty[(2)], owl:Axiom[(2)], owl:IrreflexiveProperty[(2)], owl:NegativePropertyAssertion[(2)], owl:datatypeComplementOf[(2)], owl:hasSelf[(2)]

Table 2: Survey of RDFS/OWL primitives used on the Web of Data and support in different tractable profiles where * denotes that the semantics is not fully axiomatised by the OWL RL/RDF rules or that usage of the term is restricted under OWL Direct Semantics

lion; 24%) and domains (117; 14.8%). Surprisingly (to us), RDF container membership properties (rdf:_*) are also heavily used. Regarding *prominence*, we make the following observations:

(1) The top six features are those that form the core of RDFS [28].

(2) The RDF(S) declaration classes rdfs:Class, rdf:Property are used in fewer, but more prominent documents than OWL's versions owl:Class, owl:DatatypeProperty, owl:ObjectProperty.

(3) The top eighteen features are expressible with a single RDF triple. The highest ranked primitive for which this is not the case is owl:unionOf in nineteenth position, which requires use of RDF collections (i.e., lists). Union classes are often specified as the domain or range of a given property: the most prominent such example is the SKOS vocabulary (the seventh highest ranked document) which specifies the range of the skos:member property as the union of skos:Concept and skos:Container as in (1) above.

(4) Of the features new to OWL 2, the most prominently used is owl:NamedIndividual in thirty-first position. Our crawl was conducted nineteen months after OWL 2 became a W3C Recommendation (Oct. 2009); by means of a quick scan of the max Pos column of Table 2, we note that new OWL 2 features have had little penetration in prominent Web vocabularies during that interim. Further, several OWL 2 features were not used at all in our corpus.

(5) owl:complementOf and owl:differentFrom are the least prominently used original OWL features.

In terms of profile support, we observe that RDFS has good catchment for a few of the most prominent features, but otherwise has poor coverage. Aside from syntactic/declaration features, from the top-20 features, L2 misses functional properties$_{(pos=12)}$, disjoint classes$_{(15)}$, inverse-functional properties$_{(18)}$ and union classes$_{(19)}$. RDFS-Plus omits support for disjoint$_{(15)}$ and union classes$_{(19)}$. DLP – as defined by Volz [37, §A] – has coverage of all such features, but does not support inverse-functional$_{(18)}$ datatype properties. pD* does not support disjoint$_{(15)}$ or union classes$_{(19)}$.

Regarding the OWL profiles, OWL EL and OWL QL both omit support for important top-20 features. Neither include functional$_{(12)}$ or inverse-functional properties$_{(18)}$, or union classes$_{(19)}$. OWL EL further omits support for inverse$_{(14)}$ and symmetric properties$_{(20)}$. OWL QL does not support the prevalent same-as$_{(16)}$ feature. Con-versely, OWL RL has much better coverage, albeit having only partial support for union classes(19).

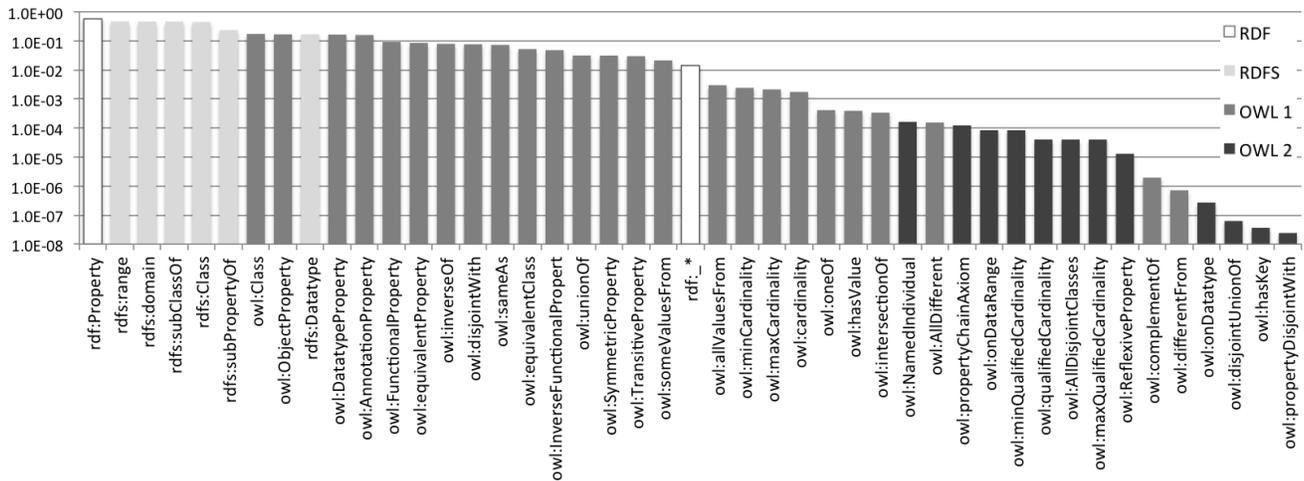

**Figure 1:** The sum of PageRank for each of the listed features from Table 2 shown in logarithmic scale on the vertical axis

Summing up, we acknowledge that such a survey of RDFS and OWL cannot give a universal or definitive indication of the most important modelling features for Linked Data. Also, OWL 2 terms might need some more time for adoption still. However, the results offer useful insights into *trends* of adoption, which inform the design of a novel OWL profile tailored for the Web of Data.

### 3.3 Survey of Datatype Use

We now look at the use of datatypes on the Web of Data.

Aside from plain literals, the RDF semantics defines a single datatype supported under RDF-entailment: rdf:XMLLiteral [17]. However, the RDF semantics also defines D-entailment, which provides interpretations over a datatype map that gives a mapping from lexical datatype strings into a value space. The datatype map may also impose disjointness constraints within its value space. These interpretations allow for determining which lexical strings are valid for a datatype, which different lexical strings refer to the same value and which to different values, and which sets of datatype values are disjoint from each other. An XSD-datatype map is then defined which extends the set of supported datatypes into those defined for XML Schema (1.0), including types for boolean, numeric, temporal, string and other forms of literals. Datatypes which are deemed to be ambiguously defined (viz. xsd:duration) or specific to XML (e.g., xsd:QName), etc. are omitted.

The original OWL specification recommends use of a similar set of datatypes to that for D-entailment, where compliant reasoners are required to support xsd:string and xsd:integer. Further, OWL allows for defining enumerated datatypes.

With the standardisation of OWL 2 came two new datatypes, namely owl:real and owl:rational. Also, OWL 2 added support for xsd:dateTimeStamp. However, XSD datatypes relating to date, time and Gregorian calendar values are not supported. Further, OWL 2 introduced mechanisms for defining new datatypes by restricting facets of legacy defined datatypes; however, we note from owl:onDatatype in Table 2 that facet restrictions have only one or two uses in our corpus.

Implementing the entire range of RDF, XSD and OWL datatypes can be costly [10], with custom code (or an external library) re-quired to support each one. Thus, it is interesting to see which datatypes are most commonly used on the Web of Data.

| # | Primitive | ∑ Rank | Lit | Doc | Dom | D | O2 |
|---|---|---|---|---|---|---|---|
| 1 | xsd:dateTime | 4.18E-2 | 2,919,518 | 1,092,048 | 68 | ✓ | ✓ |
| 2 | xsd:boolean | 2.37E-2 | 75,215 | 41,680 | 22 | ✓ | ✓ |
| 3 | xsd:integer | 1.97E-2 | 1,015,235 | 716,904 | 41 | ✓ | ✓ |
| 4 | xsd:string | 1.90E-2 | 1,629,224 | 475,397 | 76 | ✓ | ✓ |
| 5 | xsd:date | 1.82E-2 | 965,647 | 550,257 | 39 | ✓ | - |
| 6 | xsd:long | 1.63E-2 | 1,143,351 | 357,723 | 6 | ✓ | ✓ |
| 7 | xsd:anyURI | 1.61E-2 | 1,407,283 | 339,731 | 16 | ✓ | ✓ |
| 8 | xsd:int | 1.52E-2 | 2,061,837 | 400,448 | 31 | ✓ | ✓ |
| 9 | xsd:float | 9.09E-3 | 671,613 | 341,156 | 21 | ✓ | ✓ |
| 10 | xsd:gYear | 4.63E-3 | 212,887 | 159,510 | 12 | ✓ | - |
| 11 | xsd:nonNegativeInteger | 3.35E-3 | 9,230 | 10,926 | 26 | ✓ | ✓ |
| 12 | xsd:double | 2.00E-3 | 137,908 | 68,682 | 31 | ✓ | ✓ |
| 13 | xsd:decimal | 1.11E-3 | 43,747 | 13,179 | 9 | - | ✓ |
| 14 | xsd:duration | 6.99E-4 | 28,541 | 28,299 | 4 | - | - |
| 15 | xsd:gMonthDay | 5.98E-4 | 34,492 | 20,886 | 3 | ✓ | - |
| 16 | xsd:short | 5.71E-4 | 18,064 | 11,643 | 2 | ✓ | ✓ |
| 17 | rdf:XMLLiteral | 4.97E-4 | 1,580 | 791 | 11 | ✓ | ✓ |
| 18 | xsd:gMonth | 2.50E-4 | 2,250 | 1,132 | 3 | ✓ | - |
| 19 | rdf:PlainLiteral | 1.34E-4 | 109 | 19 | 2 | - | ✓ |
| 20 | xsd:gYearMonth | 8.49E-5 | 6,763 | 3,080 | 5 | ✓ | - |
| 21 | xsd:positiveInteger | 5.11E-5 | 1,423 | 1,890 | 2 | ✓ | ✓ |
| 22 | xsd:gDay | 4.26E-5 | 2,234 | 1,117 | 1 | ✓ | - |
| 23 | xsd:token | 3.56E-5 | 2,900 | 1,450 | 1 | ✓ | ✓ |
| 24 | xsd:unsignedByte | 2.62E-7 | 66 | 11 | 1 | ✓ | ✓ |
| 25 | xsd:byte | 2.60E-7 | 58 | 11 | 1 | ✓ | ✓ |
| 26 | xsd:time | 8.88E-8 | 23 | 4 | 3 | ✓ | - |
| 27 | xsd:unsignedLong | 6.71E-8 | 6 | 1 | 1 | ✓ | ✓ |
| – | *other xsd/owl dts. not used* | — | — | — | — | — | — |

**Table 3:** Survey of (std.) datatypes used on the Web of Data

In our corpus, we found 278 different datatype URIs assigned to literals. Of these, 158 came from the DBpedia exporter which models SI units, currencies, etc., as datatypes. Using analogous measures as before, Table 3 lists the top *standard* RDF(S), OWL and XSD datatypes as used to type literals in our corpus. We omit max-rank statistics for brevity, and omit plain literals which were used in 6.609 million documents (89%). D indicates the datatypes supported by D-entailment with the recommended XSD datatype map. O2 indicates the datatypes supported by OWL 2.

We observe from the table that the top four standard datatypes are supported by both the traditional XSD datatype map and in

OWL 2. However, OWL 2 does not support xsd:date$_{(5)}$ which is prominently featured in our corpus, and does not support Gregorian datatypes$_{(10,15,18,20,22)}$ nor xsd:time$_{(26)}$. Despite not being supported by any standard entailment regime, xsd:duration$_{(14)}$ was used in 28 thousand documents across four different domains.

Conversely, various standard datatypes are not used at all in the data; e.g., xsd:dateTimeStamp, the "new" OWL datatypes, binary datatypes and various normalised-string/token datatypes.

## 4. AVAILABLE TOOL SUPPORT

When asking for the practical utility of certain OWL constructs, it is crucial to consider the available tool support. In this section, we survey the availability of software that provides the necessary technical infrastructure for building complex applications, i.e., databases, reasoners and libraries.

Even if no logical inferencing is required, tools that want to support a certain OWL feature usually need to be able to read OWL documents that contain this feature or use a library for this task. Conformance with the OWL standard even requires support for the RDF/XML serialisation as an input format [34]. Parsing triples, e.g., in RDF/XML or Turtle format, into OWL axioms is not an easy task, since axioms can be composed of several RDF triples, which might be distributed all over the document [30]. In addition, OWL axioms may require use of arbitrary-length RDF lists which require particular attention to parse. Moreover, many RDF triples are ambiguous and *type declaration* axioms are necessary to resolve this. Further OWL-specific mechanisms such as imports add to the difficulty of writing an OWL parser based on one for RDF/XML or Turtle. Consequently, there are hardly any stand-alone libraries for parsing OWL (as opposed to RDF): we are only aware of the Java-based *OWL API* [21].

For tools that cannot use the OWL API due to technical or legal constraints, this puts up a major barrier for using OWL. Luckily, OWL axioms that are represented in a single RDF triple do not require the detection of complex triple patterns and can easily be processed with the RDF libraries and parsers that are available for many programming languages. The question of whether a feature can be expressed in a single triple or not may thus already have significant consequences for the practical cost of supporting it.

Databases are another important class of tools for building RDF applications and a sizeable amount of commercial and non-commercial systems is available today. Many of these systems evaluate OWL features to provide improved query answering services. Table 4 provides an overview of tools in that area. We restrict to tools that have native support at least for rdfs:subClassOf and rdfs:subPropertyOf reasoning (excluding, e.g., 5store), are developed for production use (excluding prototypes such as YARS2 [16] and QueryPie [36]) and that are meant to be used with large amounts of instance data (excluding OWL EL tools such as ELK [22]). The table lists the most frequently implemented features explicitly and describes *profile* support in a separate column. We additionally mention the main inference strategy and the source of our information.[3]

A number of tools support the (near-)complete OWL RL profile. Jena with the "OWL mini" ruleset has an incomplete implementation of OWL (1) DL features that can be viewed as an approximation of OWL RL. PelletDb and QuOnto are reasoning layers on top of a database with support for OWL DL and OWL QL, respectively.

---

[3]We note that it is difficult to verify whether the tools indeed hold what they claim, e.g., in practice one might find that the support is not as complete as advertised. Nevertheless, we take each system's description as an indication of available support.

DLEJena uses Pellet to perform TBox (schema) reasoning, where the resulting entailments and the OWL RL/RDF rules are used to generate a set of ABox (instance) rules, which are then executed using Jena's RETE engine.

Contrasting with these fairly powerful implementations, we find a number of tools that support only a few selected semantic features, including some that only support a fragment of RDFS.

The reasoning algorithms that have been used are also important in practice. Forward chaining (materialisation) often incurs significant penalties for data updates, although there are approaches to alleviate this, e.g., AllegroGraph advertises "dynamic materialisation" as a compromise. Backward chaining, in contrast, affects query answering performance but allows for easier updates. In the case of OWL QL (and RDFS), backward chaining can be performed in a particularly effective kind of query rewriting that depends on the schema information only and is thus likely to scale to bigger data volumes. The tableau approach of PelletDb, on the other hand, is more demanding when used at query time but can support all features of OWL DL.

Summarising, among the listed systems, three systems work with the Direct Semantics of OWL (PelletDb, DLEJena and QuOnto), whereas the other systems are rule-based and work directly with RDF triples, usually via forward chaining. Thus, we conclude that an implementation via rules and compatibility also with the RDF-Based Semantics is an important criteria for comprehensive tool support. Surprisingly, only two thirds of the tools support owl:sameAs, which is one of the most popular features according to our survey. A possible explanation is that owl:sameAs blows up the size of the materialisation when using forward-chaining, so for an efficient support special optimisations are required, as, e.g., implemented in OWLIM or Oracle 11g [23]. Although, four systems (nearly) support OWL RL, the complexity of a fully compliant and efficient implementation is still considered high [23].

Regarding datatypes, many triples stores use internal canonicalisation of typed literals, but full datatype reasoning is only sparsely supported or documented; some tools such as OWLIM explicitly do not support datatype rules of OWL RL. Datatype support in several tools is, for example, surveyed by Emmons et al. [10].

## 5. DEFINING THE OWL LD PROFILE

In this section, we build upon our previous observations to suggest a simple OWL profile that is adequate for the curent needs of the Web. In the previous sections, we have identified a number of key issues for OWL adoption on the Web:

1. *Adequacy*: features that are widely used on the Web should be included.
2. *Implementability*: features that are more challenging to process and reason with should be avoided.
3. *Robustness*: noisy and unreliable data should not prevent the use of ontological data in reasoning.

Comparing this to the design guidelines of RDFS-Plus [1], we can see that adequacy relates to "practicality" while implementability subsumes to "computational feasibility". We do not consider "pedagogism" as a design goal since we did not assess how intuitive features are. In contrast, the work presented in Section 3 and 4 provides us with a much better understanding for assessing implementability and adequacy. Robustness has not been considered as a design goal for RDFS-Plus while we find it to be of great importance for making sense of Web data.

Each of the above requirements leads to a number of concrete aspects. Adequacy has been discussed in Section 3 based on a sample of published ontologies. Looking at Table 2, we can see that many of the most frequently used features are of a simple structure. In fact, owl:unionOf is the highest ranked feature that is not expressed by a single triple in RDF serialisations of OWL.

| | sC | sP | ran | dom | sA | tra | sym | inv | iFP | Profile | Algorithm | Source |
|---|---|---|---|---|---|---|---|---|---|---|---|---|
| PelletDb | ✓ | ✓ | ✓ | ✓ | ✓ | ✓ | ✓ | ✓ | ✓ | OWL DL | tableau | http://clarkparsia.com/pelletdb/ |
| DLEJena | ✓ | ✓ | ✓ | ✓ | ✓ | ✓ | ✓ | ✓ | ✓ | OWL RL | tableau, forward chaining | [25], http://lpis.csd.auth.gr/systems/DLEJena/ |
| OWLIM | ✓ | ✓ | ✓ | ✓ | ✓ | ✓ | ✓ | ✓ | ✓ | ∼ OWL RL | forward chaining | [4], http://www.ontotext.com/owlim |
| Oracle 11g | ✓ | ✓ | ✓ | ✓ | ✓ | ✓ | ✓ | ✓ | ✓ | OWL RL | forward chaining | [23], http://tinyurl.com/oracle-sw |
| Jena OWL mini | ✓ | ✓ | ✓ | ✓ | ✓ | ✓ | ✓ | ✓ | ✓ | ∼ OWL RL | forward chaining | http://openjena.org/inference/ |
| Virtuoso | ✓ | ✓ | - | - | ✓ | ✓ | ✓ | ✓ | ✓ | — | backward chaining | http://virtuoso.openlinksw.com/rdf-quad-store/ |
| AllegroGraph | ✓ | ✓ | ✓ | ✓ | ✓ | ✓ | - | ✓ | - | — | forward chaining | http://tinyurl.com/agraph-doc |
| QuOnto | ✓ | ✓ | ✓ | ✓ | - | - | ✓ | ✓ | ✓ | OWL QL | query rewriting | http://www.dis.uniroma1.it/quonto/ |
| Jena RDFS | ✓ | ✓ | ✓ | ✓ | - | - | - | - | - | — | forward chaining | http://openjena.org/inference/ |
| Sesame RDFS Sail | ✓ | ✓ | ✓ | ✓ | - | - | - | - | - | — | forward chaining | http://www.openrdf.org/ |
| 4store with 4rs | ✓ | ✓ | ✓ | ✓ | - | - | - | - | - | — | query rewriting | http://4sreasoner.ecs.soton.ac.uk/ |

Table 4: RDF database systems with reasoning support (sC: rdfs:subClassOf; sP: rdfs:subPropertyOf; ran: rdfs:range; dom: rdfs:domain; sA: owl:sameAs; tra: owl:TransitiveProperty; sym: owl:SymmetricProperty; inv: owl:inverseOf; iFP: owl:InverseFunctionalProperty)

Implementability was discussed in Section 4. We observed that parsing OWL documents in RDF-based syntaxes (RDF/XML or Turtle) is easier when restricted to features that can be expressed by single triples, and which are thus directly represented in the RDF data model of available tools. Moreover, inferencing is more difficult for some features than for others, even in rule-based approaches used commonly for OWL RL, e.g., support for list-based (multi-triple) expressions that can be of arbitrary length [4].

Robustness requires a high tolerance against syntactic errors. The RDF-Based Semantics has this feature and can always be applied, hence no special language design is needed. However, it is also desirable to be able to apply the Direct Semantics to a fragment as it yields stronger completeness guarantees for reasoning. Even if RDF-Based entailments are desired, the completeness of DS reasoning methods can be used to obtain similar guarantees for RS [6, Theorem PR1]. This kind of robustness can be accomplished by reducing the use of features for which OWL DL imposes additional requirements, in particular cardinalities and property chains.

Another aspect of robustness is tolerance to inconsistencies. This feature is generally available in OWL profiles that are not able to express truly disjunctive information. Due to this, all inconsistencies are directly related to an individual or literal upon which conflicting requirements are imposed (including the special case of ill-typed literal values). Hence, it is easy to ignore (all elements involved in) inconsistencies and to continue reasoning on the remaining consistent ontology to derive meaningful conclusions. Any OWL profile (or subset thereof) has this feature.

From these observations, we derive that it is a reasonable design guideline for an OWL profile to restrict to OWL axioms that are in OWL RL and at the same time are expressed as single RDF triples. This directly addresses implementability based on the above observations together with the fact that OWL RL is now widely implemented. Adequacy is addressed since the most important features identified above are both in RL and expressed in single triples. Note that the coverage of additional, rarely used features like reflexive properties is not a concern from the viewpoint of adequacy (which asks for coverage, not for exclusivity) and is not difficult to implement in the restricted fragment either.

Robustness for interpretation in DS (i.e., as a subset of OWL DL) is eased by the omission of property chains and (most) cardinalities (note that functionality remains). However, other restrictions of OWL DL regarding the need for declarations, the non-existence of inverse functional data properties, and the restrictions on blank nodes are still relevant. We suggest to develop canonical (and thus predictable) repair strategies for addressing these issues – specifying this is left to future work. Moreover, robustness suggests that, similarly to OWL RL, arbitrary RDF graphs should be allowed when using RS for reasoning. To reconcile these issues, we first define a syntactic OWL LD profile as a subset of OWL RL (which in turn imposes the syntactic restrictions of OWL DL) and we then suggest an RS based extension of this profile for reasoning with arbitrary OWL Full ontologies.

Formally, we define OWL LD by restricting the OWL RL grammar [6]. Roughly speaking, we remove all definitions and mentions of productions listed as follows:

*Datatype entailment*:
  DataRange, DataIntersectionOf, DatatypeDefinition
*Boolean connectives & enums.*:
  *OneOf, *IntersectionOf, *UnionOf, *ComplementOf
*Restriction classes*:
  *ValuesFrom, *HasValue, zeroOrOne, *Cardinality
*Chains & keys*:
  propertyExpressionChain, HasKey
*Negative property assertions*:
  sourceIndividual, target*, Negative*PropertyAssertion

We further restrict the productions for DifferentIndividuals and Disjoint* to not use the list-based syntaxes. The full grammar can be found online [12]. All additional structural restrictions of OWL DL are inherited from OWL RL. Note that all RL datatypes are supported as well, though implementers may use our study in Section 3 to select most relevant datatypes to support (the OWL specification generally allows conforming tools to answer entailment questions with Unknown if a used feature is not supported).

Comparing OWL LD with earlier approaches, it is interesting to note that it can be viewed as a natural extension of languages like L2, RDFS-Plus, RDFS 3.0 as discussed in Section 2 and 3. In particular, RDFS 3.0 is already close to OWL LD which mainly adds further OWL 2 constructs from OWL RL while only omitting owl:AllDifferent as the list-based variant of owl:differentFrom. This adds to our confidence that OWL LD is a natural OWL profile that can be motivated from a number of perspectives.

## 6. REASONING IN OWL LD

OWL LD falls into a syntactic subset of OWL DL and can be processed by tools that implement DS entailment checking. On the other hand, we can also restrict the OWL RL/RDF rules to obtain a terse set of inference rules that yields sound but possibly incomplete entailment under RS; the full set is found in Table 5 at the end of the paper. These rules are applicable to any RDF graph allowing us to robustly draw sound conclusions from Web data.

The OWL LD ruleset comprises of rules of the form:

$$B_1 \wedge \ldots \wedge B_n \rightarrow H \; (0 \leq n \leq 3)$$

where $H$ is called the *head* and $B_1 \wedge \ldots \wedge B_n$ is the *body*. A rule with an empty body (e.g., the rule cls-thing) is simply a fact. Multiple atoms in rule heads (e.g., eq-ref) denote conjunctions that could also be expressed using multiple rules with the same body. The datatype rules are somewhat exceptional, however, and require custom logic outside of a standard rule-engine. Moreover, some rules use false in the head to express that an inconsistency is to be derived. An inconsistency-tolerant system could already be realised by simply not taking these conclusions into account for query answering.

Unlike OWL RL/RDF which encodes arbitrary-length lists in the bodies of some of its rules, the bodies of OWL LD rules comprise solely of a fixed set of (a maximum of three) ternary RDF atoms of the form $T(s, p, o)$ where $s, p, o \in \mathsf{C} \cup \mathsf{V}$. These restrictions simplify the use of the OWL LD rules in a variety of tools. Excluding datatype support, since the rules can only derive triples that are built from the set $C$ of RDF constants that originally occur in the ontology and ruleset, the number of entailments is bounded by $|C|^3$. This bound is tight, e.g., the rules entail all possible triples from the RDF graph owl:sameAs owl:sameAs a ; rdfs:domain owl:Thing . Optimisations for rule-based systems as explored in many works can be applied to implement the OWL LD inferencing efficiently. Systems can efficiently support datatypes by, e.g., only checking entailments as needed, or using canonicalisation techniques, etc.

We are now left to describe the relationship between DS and RS for the OWL LD profile.

THEOREM 1. *Let R contain the OWL LD entailment rules (Table 5) and let $O_1$ and $O_2$ be OWL 2 ontologies that satisfy the OWL LD grammar and the following properties:*

1. *neither $O_1$ nor $O_2$ contains an IRI that is used for more than one type of entity (i.e., no IRI is used both as, say, a class and an individual);*
2. *$O_1$ does not contain SubAnnotationPropertyOf, AnnotationPropertyDomain or AnnotationPropertyRange;*
3. *each axiom in $O_2$ is an assertion of the form as specified below, for $a, a_1,$ and $a_2$ named individuals:*
   (a) *ClassAssertion(C a) where C is a class,*
   (b) *ObjectPropertyAssertion(OP $a_1$ $a_2$) where OP is an object property,*
   (c) *DataPropertyAssertion(DP $a_1$ $a_2$) where DP is a data property, or*
   (d) *SameIndividual($a_1$ $a_2$).*

*Furthermore, let $RDF(O_1)$ and $RDF(O_2)$ be translations of $O_1$ and $O_2$, respectively, into RDF graphs [30]; and let $FO(RDF(O_1))$ and $FO(RDF(O_2))$ be the translation of these graphs into first-order theories in which triples are represented using the T predicate. Then, $O_1$ entails $O_2$ under the OWL 2 Direct Semantics [26] iff $FO(RDF(O_1)) \cup R$ entails $FO(RDF(O_2))$ under the standard first-order semantics.*

The proof of the Correspondence Theorem below follows immediately from the according theorem for OWL RL [6, Theorem PR1] together with the fact that OWL LD is a restriction of OWL RL. Like in the case of OWL RL, this result applies only to checking the entailment of basic facts, not of OWL axioms in general.

## 7. RELATED WORK

Here we discuss related studies on the use of the RDFS and OWL on the Web (related OWL profiles have been covered in Section 2).

One of the earliest comprehensive empirical studies of RDF Web data was presented by Ding et al. in 2005 [8]. They report about the prevalence of vocabulary terms in over 1.5 million RDF/XML Web documents, where the bulk of data was described using the Friend of a Friend (FOAF) and Dublin Core (DC) ontologies. The work focuses on characterising the structure and distributions of the raw data rather than issues relating to semantics or to RDFS and OWL.

Various works look at the syntactic profiles of OWL ontologies on the Web [2, 38, 7]. Bechhofer and Volz identify and categorise OWL DL restrictions violated by a sample group of 201 OWL ontologies (all of which were found to be in OWL Full); these include incorrect or missing typing of classes and properties, complex object properties (e.g., functional properties) declared to be transitive, inverse-functional datatype properties, and so forth [2]. In a later survey, Wang et al. study over 1,276 ontologies, where 924 (72.4%) were identified as being in OWL Full, although they proposed that 863 could be patched (93.4%) [38]. In a similar study, d'Aquin et al. found that while 81% of 22,200 RDF Web documents surveyed fell into OWL Full, from the features used, 95% would fall under the expressivity of the lightweight $\mathcal{AL}(D)$ Description Logic [7]. To summarise, these studies show that restrictions laid out in the OWL standard (specifically for the OWL Lite and OWL DL dialects) are not well-followed by Web ontologies, but that such ontologies are typically relatively inexpressive. These works re-enforce the need for our RS-based extension of OWL LD.

More recent papers focus on analysing owl:sameAs adoption on the Web of Data [9, 14]. Ding et al. provide a quantitative analysis of the owl:sameAs graph extracted from the BTC-2010 dataset (the ancestor of our corpus) [9], summarising the use of owl:sameAs to link between different publishers of Linked Data. In a similar vein, Halpin et al. [14] focus on the incorrect use of owl:sameAs [14]; they employ four human judges to manually inspect 500 such links sampled from Web data, where their results suggest that owl:sameAs is often used imprecisely, although disagreement between the judges indicates that the quality of specific owl:sameAs links can be subjective. Such surveys indicate that reasoners must proceed cautiously when operating over Web data.

## 8. CONCLUSION

We have presented a comprehensive analysis of the current use of OWL on the Web based on a large sample of RDF/XML documents. We confirmed that OWL has indeed "arrived" on the Web of Data, albeit to varying degrees for different features.

Following Linked Data principles, we used a PageRank algorithm to assess the importance of individual documents. Our results show that single-triple expressible OWL RL features are most important on the Web. A survey of existing tools confirms that these simple features tend to receive better support.

Based on these observations, we defined the OWL LD profile as a sub-language of OWL RL and provided a rule-based reasoning calculus for it. Though motivated by a new analysis of the current ontology documents on the Web of Data, OWL LD is well-aligned with the earlier proposals of RDFS-Plus and L2, indicating that it is a natural profile that can be motivated from various perspectives. We argue that this is also due to the syntactic restriction of OWL features to those that can be expressed using single RDF triples. What may appear as a superficial syntactic feature on a first glance actually identifies exactly the cases in which OWL expressions are fully aligned with the RDF data model. We argue that this bears crucial advantages regarding not only tool support but also usability. We therefore believe that, even if OWL as a whole might never arrive on the Web of Data, the OWL LD profile is a natural fit for ontological (aka. vocabulary) modelling on the Web of Data.

| | ID | Body | Head |
|---|---|---|---|
| *Equality* | EQ-REF | ?s ?p ?o . | ?s owl:sameAs ?s . ?p owl:sameAs ?p . ?o owl:sameAs ?o . |
| | EQ-SYM | ?x owl:sameAs ?y . | ?y owl:sameAs ?x . |
| | EQ-TRANS | ?x owl:sameAs ?y . ?y owl:sameAs ?z . | ?x owl:sameAs ?z . |
| | EQ-REP-S | ?s owl:sameAs ?s' . ?s ?p ?o . | ?s' ?p ?o . |
| | EQ-REP-P | ?p owl:sameAs ?p' . ?s ?p ?o . | ?s ?p' ?o . |
| | EQ-REP-O | ?o owl:sameAs ?o' . ?s ?p ?o . | ?s ?p ?o' . |
| | EQ-DIFF1 | ?x owl:sameAs ?y . ?x owl:differentFrom ?y . | false |
| *Property Axioms* | PRP-AP | (for each core annotation property ?p) | ?p a owl:AnnotationProperty . |
| | PRP-DOM | ?p rdfs:domain ?c . ?x ?p ?y . | ?x a ?c . |
| | PRP-RNG | ?p rdfs:range ?c . ?x ?p ?y . | ?y a ?c . |
| | PRP-FP | ?p a owl:FunctionalProperty . ?x ?p $?y_1$ . ?x ?p $?y_2$ . | $?y_1$ owl:sameAs $?y_2$ . |
| | PRP-IFP | ?p a owl:InverseFunctionalProperty . $?x_1$ ?p ?y . $?x_2$ ?p ?y . | $?x_1$ owl:sameAs $?x_2$ . |
| | PRP-IRP | ?p a owl:IrreflexiveProperty . ?x ?p ?x . | false |
| | PRP-SYMP | ?p a owl:SymmetricProperty . ?x ?p ?y . | ?y ?p ?x . |
| | PRP-ASYP | ?p a owl:AsymmetricProperty . ?x ?p ?y . ?y ?p ?x . | false |
| | PRP-TRP | ?p a owl:TransitiveProperty . ?x ?p ?y . ?y ?p ?z . | ?x ?p ?z . |
| | PRP-SPO1 | $?p_1$ rdfs:subPropertyOf $?p_2$ . ?x $?p_1$ ?y . | ?x $?p_2$ ?y . |
| | PRP-EQP1 | $?p_1$ owl:equivalentProperty $?p_2$ . ?x $?p_1$ ?y . | ?x $?p_2$ ?y . |
| | PRP-EQP2 | $?p_1$ owl:equivalentProperty $?p_2$ . ?x $?p_2$ ?y . | ?x $?p_1$ ?y . |
| | PRP-PDW | $?p_1$ owl:propertyDisjointWith $?p_2$ . ?x $?p_1$ ?y . ?x $?p_2$ ?y . | false |
| | PRP-INV1 | $?p_1$ owl:inverseOf $?p_2$ . ?x $?p_1$ ?y . | ?y $?p_2$ ?x . |
| | PRP-INV2 | $?p_1$ owl:inverseOf $?p_2$ . ?x $?p_2$ ?y . | ?y $?p_1$ ?x . |
| *Classes* | CLS-THING | — | owl:Thing a owl:Class . |
| | CLS-NOTHING | — | owl:Nothing a owl:Class . |
| | CLS-NOTHING2 | ?x a owl:Nothing . | false |
| *Class Ax.* | CAX-SCO | $?c_1$ rdfs:subClassOf $?c_2$ . ?x a $?c_1$ . | ?x a $?c_2$ . |
| | CAX-EQC1 | $?c_1$ owl:equivalentClass $?c_2$ . ?x a $?c_1$ . | ?x a $?c_2$ . |
| | CAX-EQC2 | $?c_1$ owl:equivalentClass $?c_2$ . ?x a $?c_2$ . | ?x a $?c_1$ . |
| | CAX-DW | $?c_1$ owl:disjointWith $?c_2$ . ?x a $?c_1$ , $?c_2$ . | false |
| *Datatypes* | DT-TYPE1 | (for each supported datatype ?dt) | ?dt a rdfs:Datatype . |
| | DT-TYPE2 | (for each literal ?lt in the value space of datatype ?dt) | ?lt a ?dt . |
| | DT-EQ | (for all $?lt_1$ and $?lt_2$ with the same data value) | $?lt_1$ owl:sameAs $?lt_2$ . |
| | DT-DIFF | (for all $?lt_1$ and $?lt_2$ with different data values) | $?lt_1$ owl:differentFrom $?lt_2$ . |
| | DT-NOT-TYPE | ?lt a ?dt . (where ?lt is not in the value space of ?dt) | false |
| *Schema Vocabulary* | SCM-CLS | ?c a owl:Class . | ?c rdfs:subClassOf ?c . ?c rdfs:subClassOf owl:Thing . ?c owl:equivalentClass ?c . owl:Nothing rdfs:subClassOf ?c . |
| | SCM-SCO | $?c_1$ rdfs:subClassOf $?c_2$ . $?c_2$ rdfs:subClassOf $?c_3$ . | $?c_1$ rdfs:subClassOf $?c_3$ . |
| | SCM-EQC1 | $?c_1$ owl:equivalentClass $?c_2$ . | $?c_1$ rdfs:subClassOf $?c_2$ . $?c_2$ rdfs:subClassOf $?c_1$ . |
| | SCM-EQC2 | $?c_1$ rdfs:subClassOf $?c_2$ . $?c_2$ rdfs:subClassOf $?c_1$ . | $?c_1$ owl:equivalentClass $?c_2$ . |
| | SCM-OP | ?p a owl:ObjectProperty . | ?p rdfs:subPropertyOf ?p . ?p owl:equivalentProperty ?p . |
| | SCM-DP | ?p a owl:DatatypeProperty . | ?p rdfs:subPropertyOf ?p . ?p owl:equivalentProperty ?p . |
| | SCM-SPO | $?p_1$ rdfs:subPropertyOf $?p_2$ . $?p_2$ rdfs:subPropertyOf $?p_3$ . | $?p_1$ rdfs:subPropertyOf $?p_3$ . |
| | SCM-EQP1 | $?p_1$ owl:equivalentProperty $?p_2$ . | $?p_1$ rdfs:subPropertyOf $?p_2$ . $?p_2$ rdfs:subPropertyOf $?p_1$ . |
| | SCM-EQP2 | $?p_1$ rdfs:subPropertyOf $?p_2$ . $?p_2$ rdfs:subPropertyOf $?p_1$ . | $?p_1$ owl:equivalentProperty $?p_2$ . |
| | SCM-DOM1 | ?p rdfs:domain $?c_1$ . $?c_1$ rdfs:subClassOf $?c_2$ . | ?p rdfs:domain $?c_2$ . |
| | SCM-DOM2 | $?p_2$ rdfs:domain ?c . $?p_1$ rdfs:subPropertyOf $?p_2$ . | $?p_1$ rdfs:domain ?c . |
| | SCM-RNG1 | ?p rdfs:range $?c_1$ . $?c_1$ rdfs:subClassOf $?c_2$ . | ?p rdfs:range $?c_2$ . |
| | SCM-RNG2 | $?p_2$ rdfs:range ?c . $?p_1$ rdfs:subPropertyOf $?p_2$ . | $?p_1$ rdfs:range ?c . |

Table 5: **The OWL LD ruleset in Turtle/N3 style syntax where** false **in the head denotes inconsistency**